\documentclass[11pt,a4paper]{article}
\usepackage{amsmath,amssymb}
\usepackage{epsfig,graphicx}

\newcommand{\be}{\begin{eqnarray}}
\newcommand{\ee}{\end{eqnarray}}
\newcommand{\bi}{\bibitem}

\def\-g{\sqrt{-g}}

\begin{document}

\title{\bf Stars older than the universe and possible mechanism of their creation }
\author{A.~D.~Dolgovr$^{a,b,c}$\footnote{{\bf e-mail}: dolgov@fe.infn.it}
\\
$^a$ \small{\em  University of Ferrara 
} \\
\small{\em Ferrara 40100, Italy}\\
$^b$ \small{\em  Novosibirsk State University
} \\
\small{\em Novosibirsk, 630090, Russia
}\\
$^c$ \small{\em  ITEP
} \\
\small{\em Moscow, 117218, Russia
}\\
}
\date{}
\maketitle

\begin{abstract}
An impressive bulk of multiple astronomical  observations indicates that there are plenty of objects in the
universe with the age which cannot be explained by the conventional theory. A model is considered which
successfully describes all these puzzling phenomena.
\end{abstract}

\section{Introduction \label{s-intro}}

Astronomical observations demonstrate the general trend, indicating that many objects in the universe were formed much 
earlier than expected  by theory and possibly so old objects are even not allowed by  the standard theory. 
Among them there are  stars in the Milky Way, older than the Galaxy and even 
older than the universe, within at least two sigma, distant high redshift (${ z\sim 10}$) objects. such as 
early galaxies, QSO/supermassive BHs, gamma-bursters, and early supenovae.

If no explanation is found in the conventional frameworks, there are two 
possible ways (maybe more) to explain these puzzling phenomena: \\
1. A novel mechanism of formation of stellar-type objects in very early universe~\cite{ad-sb}.\\
2. Modification of the cosmological expansion regime in such a way that the universe becomes older than calculated in the
frameworks of the standard model, as is done e.g. in ref.~\cite{ad-vh-it}.

Here we discuss the first possibility, more detail of which can be found in our paper~\cite{ad-sb}.
First, let us present the expression of the universe age $t_U$,
as a function of the cosmological redshift $z$:
 \be
t(z) = \frac{1}{H}\,\int_0^{{1}/({z+1)}} \frac{dx}
{\sqrt{1-\Omega_{tot} +{\Omega_m}/{x} + x^2\,\Omega_v } },
\label{t-U}
\ee 
where $\Omega_a$ is the fractional energy density of matter ($\Omega_m$, baryonic plus dark matter), of dark energy 
($\Omega_{v}$), and of the total cosmological energy density ($\Omega_{tot}$).
According to the Planck data, the present day values of theses parameters are:
${\Omega_{tot} = 1}$, ${\Omega_m  = 0.317}$, and
${\Omega_v = 0.683}$. There is some tension between the values of
the Hubble parameter measured by Planck, ${ H_{pl} = 67.3}$ km/sec/Mpc and by the traditional astronomical methods,
which can lead to $H$ as large as ${ H_{astr} =~74}$~km/sec/Mpc, see ref.~\cite{planck-prmtr} for discussion. We present a few examples
of the universe age in gigayears for different $z$, the first number corresponds to the Planck value of $H$, and the other, shorter one,
in brackets to the larger astronomical value:
$ t_U \equiv t(0) = 13.8 \,(12.5.)$; $ t(3) =  2.14\, (1.94)$; ${ t(6) = { 0.93;}\, {(0.82)}}$; ${ t(10) = 0.47\,{(0.43)}}$; and
${ t(12) = { 0.37;}\,\, {(0.33)}}$.

\section{Old stars in the Milky Way \label{s-old-stars}}

Recently several stars have been discovered in the Galaxy which ages are unexpectedly high.

Employing thorium and uranium abundances in comparison with each other and with several stable elements 
{the age of metal-poor, halo star BD+17$^o$ 3248 was estimated as} ${13.8\pm 4}$ Gyr, as is argued in
ref.~\cite{bd-17}. For comparison the estimated age of  the inner halo of the Galaxy is  ${11.4\pm 0.7}$ Gyr~\cite{halo} 

The age of a star in the galactic halo, HE 1523-0901, was estimated to be 
about 13.2 Gyr~\cite{he-1523}.
In this work  many different chronometers, such as the U/Th, U/Ir, Th/Eu, and Th/Os ratios to
measure the star age have been employed for the first time.

Most puzzling is probably the determination of the age of metal deficient {high velocity} subgiant in the solar neighborhood,
HD 140283, which has  the age ${14.46 \pm 0.31 }$ Gyr~\cite{hd-1402}.
The central value of the age exceeds the universe age by two standard deviations, if ${H= 67.3}$km/sec/Mpc, and for larger 
$H = 74$ km/sec/Mpc the star  is older than the universe more than by six standard deviations.

\section{High  redshift distant objects \label{s-hig-z} }

\subsection{Galaxies \label{ss-galaxies}}

Galaxies at high redshifts, $z \sim 10$, cannot be observed with the usual optical telescopes, which are not sensitive enough 
for such distant objects. Fortunately natural gravitational lens telescopes allow to see them, if the "telescope" happens to 
be on the light ray form the galaxy to terrestrial observers. In such a way a  galaxy at {${z \approx 9.6}$} was 
discovered~\cite{gal-500}. The galaxy was formed when the universe was about 500 million years old.

Even more striking, a galaxy at {${z \approx 11}$} has been observed~\cite{gal-300} 
which was formed before the universe age was { 0.41 Gyr} (or even shorter with larger H).

Star formation is inhibited in interstellar gas with low metallicity, because higher fraction of metals enhances 
gas cooling and facilitates gravitational capture. Interstellar medium is enriched by metals through supernova
explosion. So we need either abundant very early supervonae, or an unusual source of metallicity,
and a new mechanism of galaxy formation.
To make things more puzzling some observed early galaxies are indeed enriched with metals, see
subsection~\ref{ss-sn}.

Quoting ref.~\cite{melia}:
"Observations with WFC3/IR on the Hubble Space Telescope and the use of gravitational lensing techniques 
have facilitated the discovery of galaxies as far back as z ~ 10-12, a truly remarkable achievement. However, this 
rapid emergence of high-z galaxies, barely ~ 200 Myr after the transition from Population III star formation to Population II, 
appears to be in conflict with the standard view of how the early Universe evolved." - the quotation to be continued on top 
of the next  subsection.

\subsection{Quasars and supermassive black holes at high $z$ and now \label{ss-qso-BH}}

Continuing the quotation from ref.~\cite{melia}:
"This problem is very reminiscent of the better known (and
probably related)  premature appearance of supermassive
black holes at ${z\sim 6}$. It is difficult to understand how ${10^9 M_\odot}$ black holes
appeared so quickly after the big bang {without invoking non-standard accretion physics
and the formation of massive seeds, both of which are not seen in the local Universe."

A quasar with maximum {${ z = 7.085}$} has been discovered~\cite{qso-7}, i.e. it was 
formed at {${t< 0.75}$ Gyr.} Its luminosity is {${6.3 \cdot 10^{13} L_\odot}$} 
and mass {${2 \cdot 10^9 M_\odot}$.
{The quasars are supposed to be supermassive black holes (BH) }
{and their formation in such short time looks problematic by conventional mechanisms.}

There are strong indications that every large galaxy, as well as some relatively small ones,
contain central supermassive black hole.
The mass of the black hole may be larger than ten billions $M_\odot$ in giant elliptical
and compact lenticular galaxies and about a few million $M_\odot$ in spiral galaxies like Milky Way.
The mass of BH is typically 0.1\% of the mass of the stellar bulge of galaxy,
% \cite{BH-bulge,*SaniMarconi}
but some galaxies may  have huge BH: {e.g. NGC 1277  has
the central BH  of  ${1.7 \times 10^{10} M_\odot}$, or ${60}$\% of its bulge mass~\cite{NGC1277}.
Another interesting example is a possible existence of a supermassive black hole in an Ultracompact Dwarf Galaxy,
M60-UCD1~\cite{udg}  with the mass of about 20 million solar
mass, which is 15\% of the object's total mass. According to the conclusion of the authors, the high black hole mass and
mass fraction suggest that M60-UCD1 is the stripped nucleus of a galaxy. On the other hand, the authors
observed  "that M60-UCD1's stellar mass is consistent with its
luminosity, implying many other UCDs may also host supermassive black holes.
This suggests a substantial population of previously unnoticed supermassive
black holes."

These facts create serious problems for the standard scenario of formation of central supermassive 
BHs by accretion of matter in the central part of a galaxy.
{An inverted picture looks more plausible, when first a supermassive black holes were formed and 
attracted matter serving as seeds for subsequent galaxy formation.}

\subsection{Early Supenovae \label{ss-sn}}

{The medium around the observed early quasars contains
considerable amount of ``metals''} (elements heavier than He). 
According to the standard picture, only elements up to ${^4}$He  { and traces of Li, Be, B}
were formed in the early universe by BBN, {while heavier elements were created
by stellar nucleosynthesis and} {dispersed in the interstellar space by supernova explosions.}
{If so, prior to QSO creation a rapid star formation should take place.}
{These stars had to produce plenty of supernovae which might enrich interstellar space   by metals.}

Observations of high redshift gamma ray bursters (GBR) 
also indicate {a high abundance of supernova at large redshifts,} if GBRs
are very early supernovae.
The highest redshift of the observed GBR is 9.4~\cite{GBR-max} and there are a few more
GBRs with smaller but still high redshifts. 
{The necessary star formation rate for explanation of these early
GBRs is at odds with the canonical star formation theory.}

{A recent discovery~\cite{gal-10} of an ultra-compact dwarf galaxy
older than 10 Gyr, enriched with metals, and probably with a massive black hole in its center} 
seems to be at odds with the standard model as well.
The dynamical mass of this galaxy is ${2\times 10^8 M_\odot}$ 
and  its radius is ${R \sim 24}$ pc,  so the galaxy density is extremely high.}
There is a variable central X-ray source with luminocity{ ${L_X \sim 10^{38}}$ erg/s,} which may be
{an AGN associated with a massive black hole} or a low-mass X-ray binary.

\section{A model of formation of compact stellar-like objects and heavy PBH  in the very early universe \label{s-model-creation}}

Quite probably the described above puzzling existence of very old objects in the early high metallicity universe will find an
explanation in the frameworks of the conventional astrophysics. However, in absence of such explanation a search for 
mechanisms based on new physics is also desirable. We present here an explanation based on early works~\cite{ad-js},
where a simple generalization of the well known Affleck-Dine~\cite{ad-bs} scenario of baryogenesis allows to explain all 
observational data described above. 

The modification of the Affleck-Dine (AD) scenario of baryogenesis, which give rise to significant  production of stellar-like
objects or heavy primordial black holes can be achieved by a simple addition of  a 
general renormalizable coupling of the scalar baryon, ${\chi}$, to the inflaton field,  ${\Phi}$:
\be
U(\chi, \Phi) = U_\chi (\chi) + U_\Phi (\Phi) + U_{\rm int} (\chi,\Phi).
\label{U-of-Phi-chi}
\ee
Here $ {U_\Phi (\Phi) }$ is the inflaton potential,  ${ U_\chi (\chi)}$ is
the quartic Affleck-Dine potential, which generically has some flat directions (valleys).
The potential has the form:
\be 
U_\chi (\chi) = [m_\chi^2 \chi^2 + \lambda_\chi (\chi^4 + |\chi|^4)  + h.c.] +\lambda_2|\chi|^4\ln{\frac{|\chi|^2}{\sigma^2}},
\label{U-of-chi}
\ee 
where the last term is the Coleman-Weinberg correction~\cite{CW}, 
which arises as a result of summation of one-loop  diagrams 
in scalar field theory with quartic interaction term.

In the classical AD-scenario field ${\chi}$ acquires  a large  expectation value along a flat
direction, e.g. during inflation and evolves down later, when the Hubble parameter dropped below $m_\chi$. 
If flat directions in quadratic and quartic parts of the potential do not coincide, then at the
approach to the minimum ${\chi}$ starts to "rotate" in two dimensional $\{Re \chi, Im \chi\}$-plane.  {Rotation means that ${\chi}$
acquires (a large) average baryonic number.} 

The  additional interaction term of $\chi$ with the inflaton, $\Phi$, is  taken in the form:
\be 
U_{\rm int} (\chi, \Phi) = \lambda_1 |\chi|^2 \left( \Phi - \Phi_1\right)^2 ,
\label{U-int}
\ee
where ${\Phi_1}$ is some value of the inflaton field which it passes during inflation  and ${\lambda_1}$ is a constant.
So there is a mild tuning, but otherwise this is general renormalizable coupling between $\chi$ and $\Phi$, which
surely must exist. This terms acts as a positive time-dependent mass and thus it almost always kept the gate to the valleys closed, 
except for a short period when when ${\Phi}$ is near ${\Phi_1}$.}  
So there is a small chance for $\chi$ to reach a high value and to create a large baryon asymmetry. The behavior of the
potential $ U_\chi  (\chi) + U_{int} (\chi, \Phi)$ for different values of the effective mass is presented in fig.~\ref{fig:Potevolution}. 
The potential evolves down from the upper to the lower curve reaching the latter when $\Phi = \Phi_1$ and then the potential
returns back to the higher curve, when $\Phi$ drops below $\Phi_1$. 

\begin{figure}
	\centering
		\includegraphics[scale=0.6]{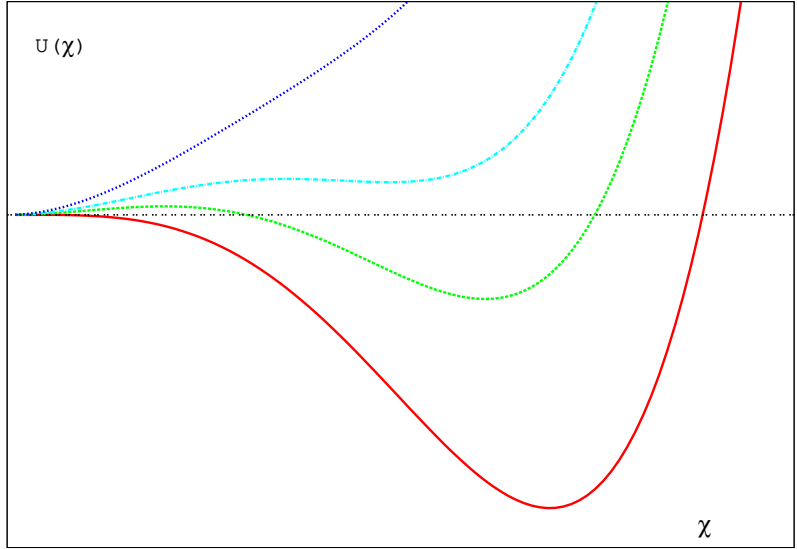}
		\caption{ {Behavior of $U_\chi(\chi)$ for different values of
    $m^2_{eff} (t)$.}}
	\label{fig:Potevolution}
\end{figure}

Correspondingly field $\chi$ rolls down toward the deeper minimum, oscillates there following the evolution of the minimum, 
rolls back to the origin, and starts to rotate around it, as is shown in fig.~\ref{fig:Chimodevol}.

\begin{figure}[htbp]
	\centering
		\includegraphics[scale=1.0]{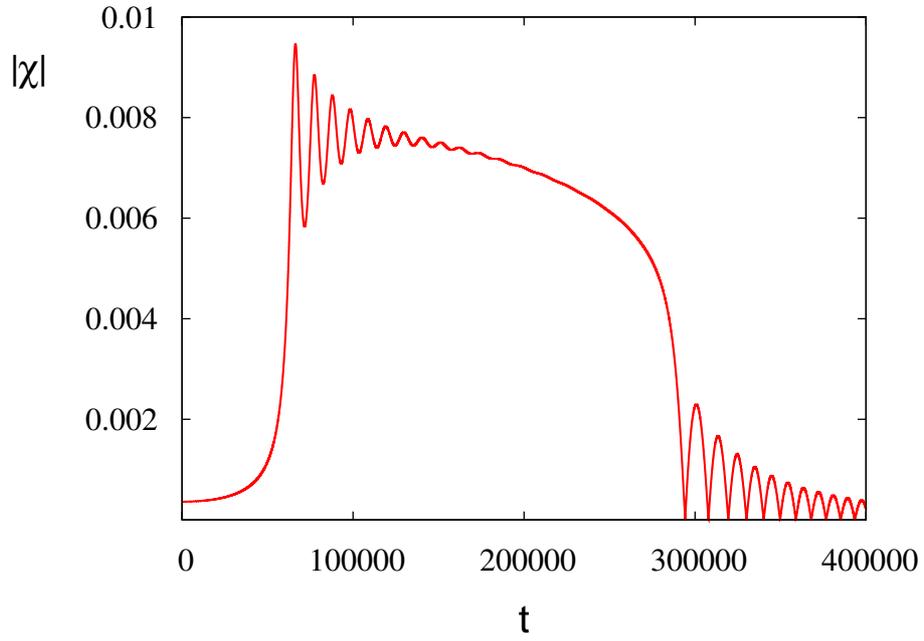}
		\caption{ Evolution of $|\chi|$ with time. }
	\label{fig:Chimodevol}
\end{figure}

Since the inflaton opens the gate to the deeper minimum only for a short time, 
the probability  for ${\chi}$ to reach the high value is low, so in most of  space baryogenesis
creates normal tiny baryon asymmetry, but {in some bubbles which occupy a small fraction of the 
whole volume, baryon asymmetry may be huge.}

After the QCD phase transition, the contrast in baryonic charge density transformed into 
perturbations  of the energy/mass density and {the bubbles with high ${B}$  formed 
PBH's or compact stellar-like objects.} The mass distribution of these high-B bubbles has
practically model independent form:
\be
\frac{dN}{dM} = C_M  \exp{ \left[-\gamma\,  \ln^2\frac{(M-M_1)^2}{ M_0^2} \right] }  \,.
\label{dN-dM}
\ee
with the model dependent parameters $C_M$, $M_1$, and $M_0$.

The values of the parameters can be adjusted in such a way that superheavy 
BHs formed at the tail of this distribution would be abundant enough to be present in every large 
galaxy and in some small ones. Such heavy PBHs could be seeds for the galaxy formation.
As we mentioned above, there is no satisfactory mechanism of creation of superheavy black holes
in the frameworks of the usual physics, while the considered here mechanism can successfully 
achieve that.

This mass distribution  naturally explains some unusual features 
of stellar mass black holes in the Galaxy.
It was found that their masses are concentrated in narrow range
${ (7.8 \pm 1.2) M_\odot }$~\cite{bh-78}.
 This result agrees with another paper where
a peak around ${8M_\odot}$, a paucity of sources with masses below
 ${5M_\odot}$, and a sharp drop-off above
${10M_\odot}$ are observed~\cite{bh-10}. {These features are not explained in the standard model.}

A modifications of ${U_{int}}$ leads to a more complicated mass spectrum of the early formed stellar type objects,
e.g.,  if:
\be
U_{\rm int}  = {\lambda_1 |\chi|^2 } 
 \left( \Phi - \Phi_1\right)^2   \left( \Phi - \Phi_2\right)^2 ,
\label{U-int-2}
\ee
we come to a two-peak mass distribution of the PBHs and compact stars, which is 
probably observed~\cite{bh-two-peak}, but not yet explained. 

{Evolved chemistry in the so early formed QSOs can be explained, at least to some extend,}
{by more efficient production of metals during BBN due to much larger ratio 
${\beta =N_B/ N_\gamma}$. }
The standard BBN essentially stops at $^4$He due to very
small ${\beta}$. However, in the model considered here {${\beta}$ is
much larger than the canonical value, even being close or exceeding unity.} In such conditions
much heavier primordial elements can be produced~\cite{jap-bbn}. It is possible that stars which
initiated with more metals than the usual ones could look older than they are and if their age is
evaluated by the standard nuclear chronology, they might even look older than the universe.

\section{Conclusion \label{ss-concl}}

{The scenario may be speculative but not too unnatural and explains a lot:} \\
1. Superheavy BH and early quasar formation with plenty of metals around.\\
{2. High abundance of supenovae and gamma-bursters at ${z\gg 1}$.}\\
{3. Existence very old stars in the Galaxy} and very old galaxies.\\[2mm]
{Additionally, new types of stellar-like objects from the very early universe  and 
probably abundant cosmic antimatter in the Galaxy are predicted~\cite{bambi-ad}. 
A study of astrophysics of such new kind of stars is in order.\\[3mm]
{\bf Acknowledgement.} This work was  supported by the grant of the Russian Federation government
11.G34.31.0047.

\end{document}